\documentclass[sigconf,nonacm]{acmart}

\usepackage{tabularx}
\usepackage{placeins}
\usepackage{acronym}
\usepackage{amsmath}
\usepackage{booktabs}
\usepackage{float}
\renewcommand\footnotetextcopyrightpermission[1]{}

\acrodef{snn}[SNN]{Spiking Neural Network}
\acrodef{reram}[ReRAM]{Resistive random-access memory}
\acrodef{mi}[MI]{Model Inversion}
\acrodef{sdmi}[SDMI]{Spiking Domain Model Inversion}
\acrodef{eot}[EOT]{Expectation over Transformation}
\acrodef{adc}[ADC]{Analog-to-Digital Converter}
\acrodef{dac}[DAC]{Digital-to-Analog Converter}

\begin{document}


\title{Hardware-Aware Model Inversion in Memristive Spiking Neural Networks}

\author{Pablo Sorrentino}
\affiliation{%
  \institution{University of Groningen}
  \city{Groningen}
  \country{The Netherlands}
}
\email{p.f.a.sorrentino@rug.nl}

\author{Fernando M. Quintana}
\authornote{Present affiliation: Technical University of Munich, Munich, Germany.}
\affiliation{%
  \institution{University of Groningen}
  \city{Groningen}
  \country{The Netherlands}
}
\email{fernando.quintana@tum.de}

\author{Erika Covi}
\authornotemark[1]
\affiliation{%
  \institution{University of Groningen}
  \city{Groningen}
  \country{The Netherlands}
}
\email{erika.covi@tum.de}

\author{Fatih Turkmen}
\affiliation{%
  \institution{University of Groningen}
  \city{Groningen}
  \country{The Netherlands}
}
\email{f.turkmen@rug.nl}

\renewcommand{\shortauthors}{Sorrentino et al.}

\begin{abstract}
Hardware-accelerated spiking neural networks (SNNs) implemented on memristive crossbar arrays offer significant gains in energy efficiency, while device variability and analog noise are sometimes assumed to hinder information extraction. However, whether these hardware effects improve, preserve, or degrade privacy after deployment remains largely unexplored. Therefore, we ask whether model-inversion leakage characterized on an SNN remains representative after the same model is mapped to a non-ideal memristive substrate, and how hardware non-idealities and reliability faults affecting synapses, neuron thresholds, and spike processing reshape that leakage. To study this problem, we introduce Spiking Domain Model Inversion (SDMI), an inversion method that combines Expectation over Transformation (EOT) with surrogate gradients to optimize inputs through stochastic analog behavior and non-differentiable spiking dynamics. We use SDMI consistently across the SNN model, its memristive SNN, and the same memristive SNN under additional hardware perturbations.

Across four datasets, reconstruction leakage does not vary monotonically with model accuracy or hardware variability. At comparable accuracy levels, neuron-threshold perturbations increase reconstruction fidelity, measured by SSIM, by up to 51.8\%, whereas spike-level perturbations primarily degrade utility without consistently improving reconstruction. Weight noise, in contrast, reduces leakage under the same inversion procedure. These results show that hardware variability is not inherently privacy-preserving: model-inversion leakage in memristive SNNs depends on the hardware state and cannot be inferred from model accuracy alone.
\end{abstract}


\maketitle


\section{Introduction}
Emerging in-memory and neuromorphic computing systems, particularly those based on memristive crossbar arrays, are becoming a key paradigm for energy-efficient edge intelligence \cite{ielmini2018memory, seekings2025integrated}. By co-locating computation and memory, and leveraging event-driven spiking dynamics, these systems significantly reduce data movement and energy consumption and can achieve orders-of-magnitude improvements in energy efficiency compared to conventional von Neumann architectures \cite{ielmini2020emerging, peng2024memristor} when implementing data-intensive applications.

However, the same physical properties that enable high energy efficiency and parallel computation also introduce new security and reliability challenges \cite{sorrentino2026neuromorphic,nazari2024securing,wang2024safe}. Memristive hardware inherently exhibits non-deterministic behavior due to device variability, analog noise, conductance drift, and temporal dynamics \cite{le2023using, lammie2025inherent}. These effects complicate whether functional behavior and privacy exposure characterized at the model level remain representative after deployment onto the analog substrate. Consequently, as these systems move toward real-world deployment, their privacy implications remain insufficiently understood, particularly when hardware behavior may differ from that of the software SNN.

In memristive neuromorphic systems, synaptic weights learned during training are directly mapped to device conductance states \cite{ielmini2018memory}. As a result, the trained SNN becomes tightly coupled to its physical implementation. Device variability, temporal drift, peripheral quantization, and other implementation effects can therefore alter the effective computation performed during inference \cite{ielmini2020emerging}. Prior work has further shown that neuromorphic and in-memory computing systems expose hardware-dependent security surfaces, including side-channel and fault-related behaviors \cite{wang2024cimsat, inglese2023side, sapui2024power, ensan2021scare}. These observations motivate privacy analysis that treats the hardware state as part of the deployed system rather than assuming that the software SNN fully describes inference behavior.

Despite the increasing adoption of memristive neuromorphic systems, particularly for implementing spiking neural networks (SNNs), analysis of privacy remains largely disconnected from the underlying hardware \cite{moshruba2025neuromorphic}. Existing privacy evaluations of \acp{snn} predominantly consider idealized settings, without explicitly accounting for stochastic and time-dependent effects introduced by memristive hardware \cite{poursiami2025spikes}. This raises a fundamental question: \emph{Does the privacy exposure measured on a SNN remain representative after the same model is mapped onto a non-ideal memristive substrate?} Most existing work on privacy attacks in this context is tailored to conventional neural networks and does not explicitly account for the physical substrate \cite{fredrikson2015model, shokri2017membership, wang2021variational}.

At the same time, hardware reliability studies show that deviations in synaptic storage, neuron behavior, and signal propagation can substantially alter neural inference \cite{bi2023ad2vncs}. In particular, it is unclear whether different hardware states merely degrade model utility or whether they can also reshape the susceptibility of the deployed model to data-extraction attacks. This creates a blind spot in the privacy assessment of memristive \acp{snn}: evaluating only the software SNN may overlook hardware-dependent changes that emerge after hardware mapping or under reliability faults.

\subsection{Motivation and Problem Statement}

Memristive neuromorphic systems may operate under a range of hardware conditions arising from intrinsic device non-idealities, aging, environmental disturbances, or reliability failures \cite{joksas2025nonideality, lammie2025inherent}. In this context, device-level variability in \ac{reram} crossbars becomes a double-edged property. Although such variability can support and enhance certain security primitives \cite{cambou2016puf, xu2023embedding}, it also changes the behavior of a trained model after its parameters are mapped to physical conductance states. Reliability faults can further alter the hardware state at synaptic, neuronal, or spike-processing levels. Importantly, these mechanisms are not assumed to be equally controllable by a physical adversary; they are used to characterize how different reliability regimes affect model-inversion leakage.

Furthermore, device-to-device variability, cycle-to-cycle noise, conductance drift, and finite-precision conversion introduce stochasticity that is absent from the SNN. As a result, gradients and inversion behavior computed from the model need not fully characterize inversion after memristive weight mapping. This motivates a hardware-aware analysis of model inversion under both non-differentiable spiking dynamics and stochastic analog behavior.

Existing privacy attacks on neural networks are largely developed for settings in which model outputs and internal computations are deterministic or sufficiently repeatable \cite{salem2018ml, fredrikson2015model, shokri2017membership}. In particular, many optimization-based \ac{mi} methods rely on stable confidence or logit information and differentiable or surrogate-differentiable model behavior \cite{zhu2019deep, wang2021variational}. These assumptions become more difficult to apply directly to memristive \acp{snn}, where computation is sparse, event-driven, temporally encoded, and affected by hardware stochasticity \cite{moshruba2025neuromorphic, li2025membership, lammie2025inherent, joksas2025nonideality}. Recent work has also shown that privacy exposure can vary across model architectures even at comparable predictive performance \cite{zhang2024does}.

\begin{figure}[ht]
  \centering
  \includegraphics[width=1.00\linewidth]{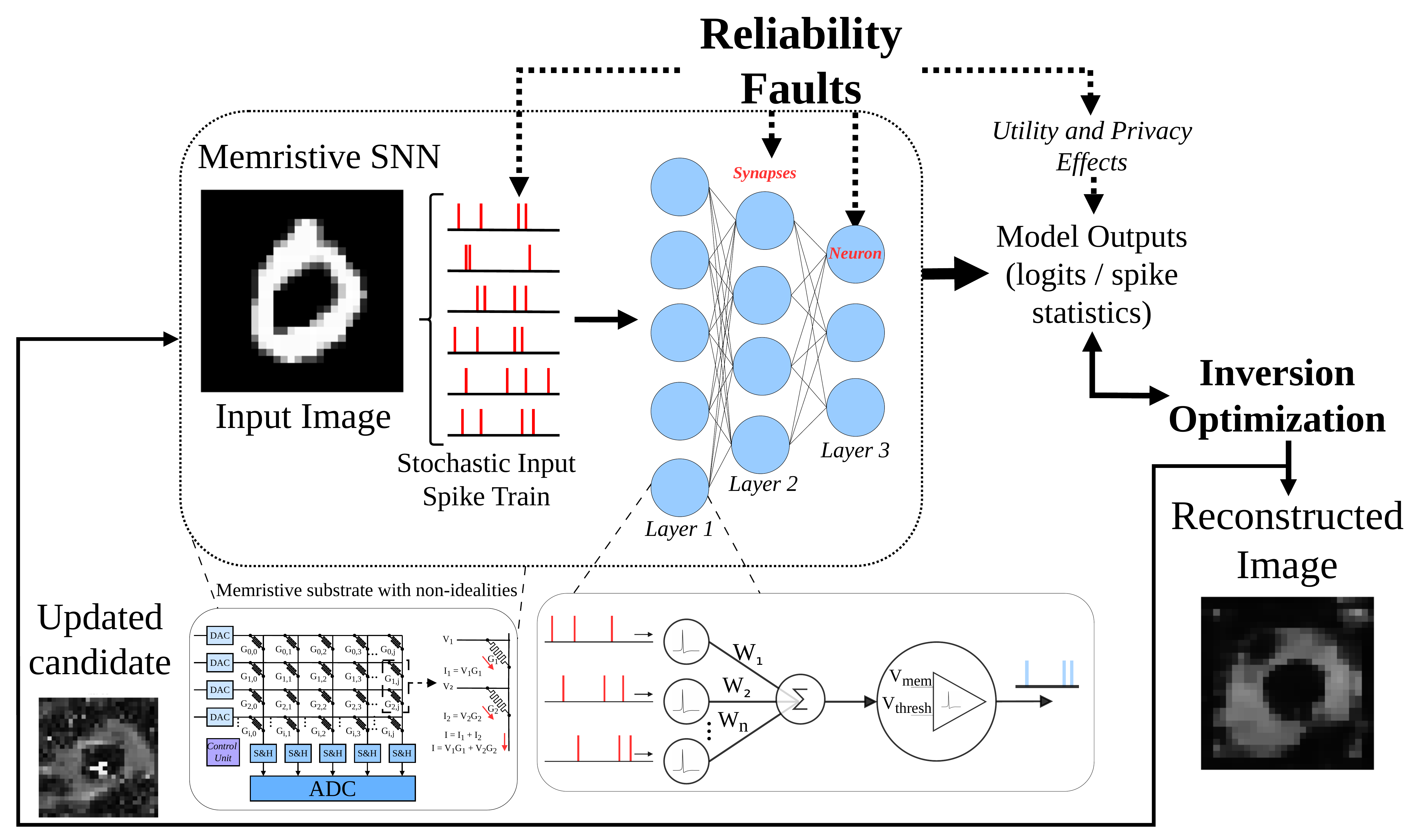}
  \caption{ Overview of model inversion in memristive SNNs. Hardware non-idealities and reliability faults can alter the effective hardware state and observable model behavior, potentially affecting both predictive utility and model-inversion leakage. Model inversion iteratively optimizes a candidate input based on model responses. }
  \label{fig:overview}
\end{figure}

This leads to two related questions. First, does model inversion performed on the SNN provide an accurate characterization of the privacy exposure of its memristive implementation? Second, once deployed, how do hardware non-idealities and reliability faults reshape the relationship between model utility and inversion leakage? We address these questions by studying Spiking Domain Model Inversion across different execution conditions of memristive \acp{snn}. Our goal is to characterize how the effective hardware state influences model-inversion leakage, rather than to demonstrate a chip-specific physical fault-injection mechanism. Figure~\ref{fig:overview} illustrates the general model inversion setting considered in this work and the presence of hardware non-idealities and reliability faults affecting the behavior of a memristive SNN.

\subsection{Contributions}

This work investigates model-inversion leakage in memristive \acp{snn} as a property of both the model and its hardware state. In summary, this paper makes the following contributions:

\begin{itemize}

\item We introduce Spiking Domain Model Inversion (SDMI), an inversion method for non-ideal memristive SNNs that combines Expectation over Transformation (EOT) with surrogate gradients to reconstruct class-representative inputs under stochastic analog behavior and non-differentiable spiking dynamics.

\item We distinguish native hardware non-idealities, reliability faults, and controlled sensitivity conditions, and characterize how deviations at synaptic, neuronal, and spike-processing levels reshape inversion behavior in memristive \acp{snn}.

\item We implement the proposed framework using a ReRAM hardware model based on IBM AIHWKIT modeling abstractions \cite{le2023using} and evaluate it on static, biometric, and event-based benchmarks, including MNIST, Fashion-MNIST, AT\&T Faces, and DVS128 Gesture.

\item Finally, we show that model utility and reconstruction leakage do not necessarily evolve together. In particular, neuron-level perturbations increase reconstructed-image similarity by up to 51.8\% relative to the memristive baseline, whereas spike-level perturbations can produce comparable accuracy degradation without corresponding gains in reconstruction similarity.

\end{itemize}

\section{Background}

In this section, we provide background on spiking neural networks, memristive crossbar inference, hardware non-idealities and reliability faults, and model inversion attacks under non-ideal neuromorphic systems.

\subsection{Spiking Neural Networks}

\acp{snn} define a paradigm shift from continuous-valued activations to discrete, temporal spike events $S \in \{0,1\}$ \cite{eshraghian2023training, neftci2019surrogate}. Unlike artificial neural networks (ANNs), inference in \acp{snn} is a dynamical process where information is encoded in the timing and rate of spike events. The internal state of a neuron is defined by its membrane potential $V[t]$, which integrates synaptic inputs over an inference window $T$. We formalize this behavior using the Leaky Integrate-and-Fire (LIF) model:

\begin{equation}
\label{eq:lifform}
V_j[t] =
\beta V_j[t-1]
+
\sum_i W_{i,j} S_i[t]
-
V_{th} S_j[t-1],
\end{equation}

where $V_j[t]$ is the membrane potential of the neuron $j$ at time $t$, $W_{i,j}$ is the synaptic weight from the presynaptic neuron $i$ to the neuron $j$, $\beta$ is the decay constant, and $V_{th}$ is the firing threshold. This temporal dependency implies that perturbations in spike timing, arising from stochastic encoding, hardware variability, or timing variations, can propagate through the network and alter firing patterns \cite{nomura2022robustness, yadav2024unveiling}. The discrete output $S_j[t]$ is defined by the Heaviside step function $H(\cdot)$:

\begin{equation}
S_j[t] = H(V_j[t] - V_{th}).
\end{equation}

Although $S_j[t]$ is discrete, the underlying membrane trajectory $V[t]$ remains sensitive to accumulated synaptic input. Perturbations that alter threshold crossings, synaptic integration, or spike propagation can therefore change the temporal output behavior.

The non-differentiability of $H(\cdot)$ prevents direct gradient-based optimization through spike generation. As described in Section~\ref{sec:privacy-attacks}, SDMI addresses this issue using surrogate gradients during the backward pass \cite{neftci2019surrogate, eshraghian2023training}.

\subsection{SNNs with Memristive Crossbars}

\begin{figure}[t]
  \centering
  \includegraphics[width=1.00\linewidth]{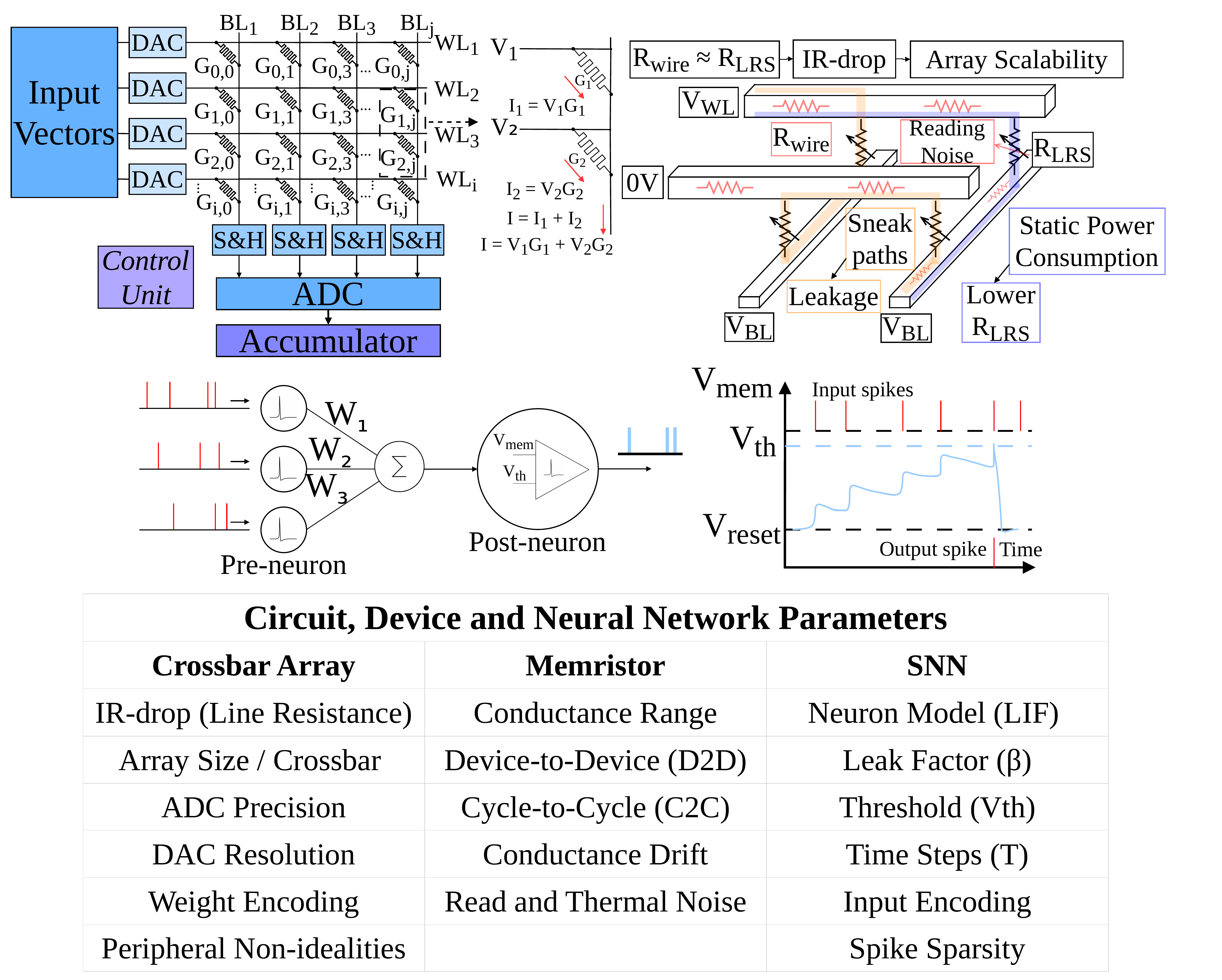}
  \caption{Hierarchical overview of representative system parameters across three levels: memristor array (IR-drop and peripheral circuitry), device (conductance variability, drift, and noise), and SNN model (neural dynamics and temporal encoding).}
  \label{fig:system_params}
\end{figure}

Synaptic weights of SNNs can be realized as the conductance states of memristive devices arranged in a crossbar topology \cite{ielmini2018memory, chen2017neurosim+, wang2017memristors, woo2022probabilistic}. Figure~\ref{fig:system_params} summarizes representative parameters across the memristive-device, crossbar, and SNN execution stack.

According to Ohm's Law and Kirchhoff's Current Law (KCL), the current $I_j$ at column $j$ is obtained by summing the currents contributed by the devices connected to that column:

\begin{equation}
I_j[t]
=
\sum_i
\left(
V_i^{in}[t] - \Delta V_{i,j}
\right)
G_{eff,i,j},
\end{equation}

where $V_i^{in}[t]$ is the input voltage applied to row $i$, $G_{eff,i,j}$ denotes the effective conductance of the device connecting row $i$ to column $j$, and $\Delta V_{i,j}$ represents attenuation of the input voltage due to \textit{IR-drop} along the interconnects \cite{brivio2022hfo2}. To support signed synaptic weights, $W_{i,j}$ is represented using a differential pair of conductances:

\begin{equation}
W_{i,j}
=
\kappa
\left(
G^{+}_{i,j} - G^{-}_{i,j}
\right),
\end{equation}

where $\kappa$ denotes the scaling factor between the conductance difference and the corresponding algorithmic weight.

The effective conductance may differ from its nominal programmed value because of device variability, conductance drift, and read-time non-idealities. These effects alter the relationship between the nominal weights and the effective computation performed during inference. Consequently, models sharing the same parameters may behave differently after mapping to a non-ideal analog substrate. This distinction between the nominal configuration and effective hardware execution state is central to our work.

The analog crossbar interfaces with surrounding logic through \acp{dac}, \acp{adc}, and sensing circuitry, introducing finite precision, quantization, saturation, and sensing effects \cite{le2023using}. Together with device-level variability and temporal conductance dynamics, these effects distinguish memristive inference from inference on the software SNN.

\subsection{Hardware Non-Idealities and Reliability Faults}
\label{subsec:hardware-perturbations}

Memristive \acp{snn} can deviate from their software behavior due to three broad causes: intrinsic device non-idealities, reliability faults in processing elements, and disturbances affecting spike communication. Similar behavioral effects may also arise under physical fault injection, although their controllability and physical cost depend on the hardware platform and perturbation mechanism.

We distinguish the physical origin of a hardware deviation from the experimental probe used to study its effect. \emph{Native hardware non-idealities}, such as conductance variability and drift, arise during ordinary analog behavior. \emph{Reliability faults} represent deviations associated with device, circuit, peripheral, or communication failures. Finally, \emph{controlled sensitivity probes} deliberately vary execution parameters in simulation to isolate their effect on model behavior and model-inversion leakage. For analysis, we organize these deviations into three functional perturbation levels: synaptic, neuron, and spike.

\subsubsection{Synaptic-Level Deviations}

Synaptic-level deviations modify the effective strength of connections represented by the memristive crossbar. They include native effects such as conductance drift and variability, as well as hardware deviations such as weight noise or errors in the weight representation associated with the mapping or peripheral logic.

\subsubsection{Neuron-Level Deviations}

Neuron-level deviations affect the dynamics of the LIF neuron, particularly its firing threshold $V_{th}$ \cite{raptis2025input}. Changes in comparator behavior, bias conditions, circuit variability, or other hardware faults can be abstracted as a deviation $\delta_{th}$ in the firing condition of the spikes:

\begin{equation}
S_j[t]
=
H
\left(
V_j[t]
-
\left(
V_{th}+\delta_{th}
\right)
\right).
\end{equation}

A negative threshold deviation increases neuronal excitability, whereas a positive deviation suppresses firing activity. Such changes modify the temporal sparsity and activation structure of the network and are therefore relevant for the connection between the model inversion attack and the effective hardware state.

\subsubsection{Spike-Level Deviations}

Spike-level deviations affect the presence or timing of discrete events transmitted between processing elements and include spike deletion, insertion, and temporal displacement \cite{colucci2022enpheeph, gougebakan2024spikingjet, yu2026time}. These deviations modify the temporal information propagated through the network and can arise from reliability- or communication-level disturbances.

\subsection{Model Inversion Attack in Spiking Neural Networks}
\label{subsec:sdmi}

Model Inversion (MI) attacks seek representative inputs or sensitive features associated with a target model output \cite{fredrikson2015model, wang2021variational}. Applying optimization-based MI to memristive \acp{snn} introduces two challenges: spike generation is discrete and temporally dependent, while hardware mapping introduces variability that is not represented by the software SNN.

We denote the trained SNN parameters by $\Theta$, the hardware state by $h$, and an optional hardware perturbation or sensitivity setting by $p$. Throughout the paper, $f_{\Theta}^{digital}$ denotes the software SNN, $f_{\Theta,h}^{mem}$ the same SNN mapped to the memristive model under hardware state $h$, and $f_{\Theta,h,p}^{mem}$ the mapped SNN under an additional perturbation $p$.

SDMI combines surrogate gradients with Expectation over Transformation (EOT) to handle discrete spiking dynamics and average over stochastic hardware and spike realizations. Section~\ref{sec:privacy-attacks} presents the complete inversion objective and optimization procedure.

\section{Related Work}

Prior work relevant to this paper spans three mostly separate research threads:
(i) privacy attacks on neural networks, primarily developed for conventional NNs without explicitly accounting for the hardware substrate;
(ii) fault injection and reliability studies, largely developed for ML accelerators and emerging computing hardware with a focus on integrity, robustness, and functional degradation; and
(iii) emerging security analyses of \acp{snn} and in-memory computing hardware, which consider temporal neural dynamics and hardware-dependent attack surfaces.
We compare these threads along five dimensions: privacy analysis, hardware perturbation modeling, temporal spiking dynamics, hardware-level execution, and their joint consideration.
Their intersection remains comparatively unexplored, particularly how reconstruction changes when temporally encoded inference is mapped onto non-ideal hardware and exposed to reliability faults.

\subsection{Privacy Attacks on Neural Networks}

Privacy attacks on machine learning models have been extensively studied in conventional neural networks, with \acf{mi} providing a mechanism for reconstructing sensitive information from trained models. Early work demonstrated that input features could be recovered by exploiting model outputs such as confidence scores and class probabilities \cite{fredrikson2015model}. Subsequent approaches improved the fidelity of reconstruction through gradient-based optimization and more expressive priors, enabling attacks in both white-box and black-box settings \cite{zhu2019deep, wang2021variational, usynin2023beyond}. More recent work has advanced inversion capabilities using generative models and latent-space reconstruction, achieving higher reconstruction fidelity and improved scalability \cite{tao2022mirror, shi2023scale}. Inversion-based attacks have also expanded beyond full input reconstruction to settings including attribute inference and leakage through model updates or unlearning mechanisms \cite{hu2024learn}.

A related line of work studies membership inference attacks, which determine whether a sample was included in the training dataset \cite{shokri2017membership, salem2018ml}. Unlike model inversion, these attacks focus on membership signals rather than reconstructing the underlying input \cite{liu2022ml}. Despite extensive studies, privacy attacks on neural networks largely omit the hardware substrate. In memristive \acp{snn}, however, hardware variability and temporal spiking dynamics can alter the model behavior after hardware mapping, motivating an analysis of whether model-level privacy characterization remains representative after hardware mapping.

\subsection{Fault Injection and Reliability in Hardware}

Prior work on neural networks has shown that faults affecting model parameters or intermediate computations can compromise model integrity. For example, bit-flip attacks on weights can significantly alter model behavior with small numbers of discrete modifications \cite{he2020defending, dong2023one}. Fault-injection studies on neural accelerators have similarly investigated timing, voltage, and memory perturbations as adversarial mechanisms and as tools for evaluating robustness under non-ideal hardware behavior \cite{hong2019terminal, wang2023nntesting}.

More recent work has explored fault behavior and reliability in compute-in-memory (CIM) and memristive architectures, where device-level non-idealities, mixed-signal operation, and peripheral circuitry introduce additional sources of hardware deviation \cite{inglese2023side, sapui2024power, wang2024cimsat}. Studies have also analyzed fault resilience in neuromorphic and memristive systems, including device faults in spiking neural networks trained with biologically inspired rules such as spike-timing-dependent plasticity \cite{jouni2025stdp}, as well as fault behavior in memristor-based logic-in-memory implementations and spiking neural networks \cite{staudigl2022x, oli2022stuck, gougebakan2024spikingjet}.

This body of work establishes that deviations at synaptic, neuron, peripheral, and communication levels can materially change neural behavior. From a security perspective, fault injection provides an adversarial mechanism for intentionally inducing a subset of such deviations, thereby connecting reliability analysis with attacker manipulation without implying equivalent physical controllability across fault classes. However, these studies predominantly evaluate functional correctness, robustness, or integrity rather than whether different hardware states reshape susceptibility to information-extraction attacks. We use this reliability and fault literature as the basis for our privacy characterization.

\subsection{Security of SNNs and Neuromorphic Systems}

Security research on \acp{snn} neuromorphic platforms has gained attention only recently, spanning hardware-level attacks, software-level threats, and emerging privacy concerns \cite{sorrentino2026neuromorphic}. Prior studies have shown that spiking neural networks are vulnerable to adversarial examples adapted to spike-based encoding schemes, as well as to backdoor attacks embedded in temporal spike patterns \cite{abad2023sneaky, nomura2022robustness, yadav2024unveiling}. These works demonstrate that \acp{snn} are not inherently robust to adversarial manipulation despite their event-driven nature.

A separate line of work has begun to explore privacy risks in neuromorphic systems. Recent studies investigate whether neuromorphic architectures provide inherent privacy benefits and evaluate model inversion attacks in spiking settings \cite{poursiami2024brainleaks, poursiami2025spikes}. Prior SNN inversion methods address discrete spiking inputs through surrogate-gradient or distribution-based optimization, while recent black-box approaches study inversion through surrogate modeling. Neither considers whether memristive mapping and hardware-state variability reshape inversion leakage for the same trained SNN.

In parallel, hardware-oriented studies have analyzed side-channel leakage and fault-related vulnerabilities in neuromorphic and in-memory computing systems \cite{wang2024cimsat, goswami2023experimental, wang2023side, probst2024side}. These works show that device-level behavior and hardware dynamics can create observable hardware-dependent security effects, but do not systematically connect non-ideal hardware behavior to information-extraction attacks such as model inversion.

A related line of work studies privacy- and security-aware design in memristive and compute-in-memory hardware. Prior works examine adversarial robustness, device-variation tolerance, and secure CIM design \cite{wang2024safe, bi2023ad2vncs, lammie2025inherent}, while others explore hardware-assisted privacy-preserving learning through memristive implementations of differential privacy \cite{fu2019memristor, fu2020memristor}. Protection mechanisms specific to neuromorphic systems, such as watermarking for \acp{snn}, have also been proposed \cite{poursiami2024watermarking}. These approaches demonstrate that hardware characteristics can influence security and privacy properties, but they do not quantify how memristive non-idealities and hardware perturbations at the synaptic, neuron, and spike levels affect model-inversion leakage in \acp{snn}.

As summarized in Table~\ref{tab:comparison}, most existing works consider information extraction, hardware perturbations, or temporal neuromorphic security separately, without jointly considering their intersection. Our work connects these dimensions by studying the reconstruction across software and memristive \acp{snn} and under different hardware perturbations.

\begin{table*}[t]
\centering
\small
\setlength{\tabcolsep}{3pt}
\caption{Comparison with prior work on privacy attacks, hardware perturbations, and neuromorphic security.}
\label{tab:comparison}
\begin{tabular}{lcccccc}
\hline
\textbf{Prior Work} & \textbf{Target System} & \textbf{Privacy Analysis} & \textbf{Reliability} & \textbf{Temporal SNN} & \textbf{Hardware} & \textbf{Privacy and HW} \\
\hline

ANN MI/MIA \cite{fredrikson2015model, shokri2017membership} 
& ANN & Yes & No & No & No & No \\

Non-ideality-aware ML \cite{lammie2025inherent, joksas2025nonideality} 
& ANN & No & Yes & No & Yes & No \\

DNN fault injection \cite{he2020defending, hong2019terminal} 
& ANN/DNN & No & Yes & No & Yes & No \\

FIA/SCA on CIM \cite{sapui2024power, wang2024cimsat} 
& ANN/CIM & No & Yes & No & Yes & No \\

SNN security \cite{abad2023sneaky, yadav2024unveiling} 
& SNN & No & No & Yes & No & No \\

SNN privacy \cite{moshruba2025neuromorphic, poursiami2025spikes} 
& SNN & Yes & No & Yes & No & No \\

Neuromorphic leakage \cite{goswami2023experimental, probst2024side} 
& SNN/IMC & No & Yes & Yes & Yes & No \\

\textbf{This Work} 
& \textbf{Memr. SNN} & \textbf{Yes} & \textbf{Yes} & \textbf{Yes} & \textbf{Yes} & \textbf{Yes} \\

\hline
\end{tabular}
\end{table*}

\section{Threat Model}
\label{sec:threatmodel}

We assume that the adversary's goal is to extract information from a trained target \ac{snn}. The privacy attack considered is SDMI (Section~\ref{sec:privacy-attacks}), which reconstructs inputs representative of a target class by optimizing a candidate input toward a target model output \cite{poursiami2025spikes}. Throughout the paper, reconstruction leakage refers to class-dependent information recoverable through model inversion; it does not imply exact recovery of an individual training record. We do not consider model accuracy degradation or compromise of model integrity as the adversary's objective \cite{wen2025sok}. 

\subsection{Adversary Knowledge and Observability}

We consider a white-box adversary with access to the trained \ac{snn}, including its architecture and trained parameters, and knowledge of the input encoding procedure. We intentionally adopt this setting to isolate hardware-dependent privacy effects from uncertainty introduced by SNN model extraction or surrogate mismatch. However, knowledge of the trained model does not imply direct physical knowledge of the effective analog hardware state. After mapping, inference depends on a hardware state $h$ that captures the conductance configuration and stochastic hardware effects represented in our evaluation. The adversary does not obtain direct readout of effective conductances, instantaneous device variability, internal membrane potentials $V[t]$, individual device states, or gradients from the deployed hardware.

For each evaluated hardware state, the evaluator instantiates the corresponding differentiable model to measure inversion susceptibility. This provides a comparative privacy characterization and does not assume that a deployed attacker can infer the instantaneous hardware state or reproduce the evaluated perturbation. SDMI optimizes the time-averaged output logits, while EOT averages over stochastic hardware and spike samples. The hardware state is therefore part of the target being characterized rather than an attacker-controlled input to SDMI.

Importantly, direct observation of the hardware state, on-device gradients, or precise fault control are not required to perform model inversion given knowledge of the trained \ac{snn}. Such capabilities could nevertheless strengthen an attacker by reducing the mismatch between the trained model and its physical realization, or by steering the hardware toward states with greater inversion leakage; we do not assume these capabilities. Our question is therefore whether the privacy characterization obtained from the trained model remains representative after mapping to a non-ideal memristive substrate.

\subsection{Hardware States and Faults}

The adversary controls the inversion procedure, not the occurrence of every evaluated hardware perturbation. We consider hardware deviations at synaptic, neuronal, and spike processing levels. As described in Section~\ref{subsec:hardware-perturbations}, these conditions include native hardware non-idealities, reliability faults, and controlled sensitivity probes; they are not assumed to be equally controllable by an adversary.

Here, reliability refers to deviations of the deployed memristive \ac{snn} from its intended hardware behavior due to device, circuit, peripheral, or communication effects. Native hardware non-idealities require no attacker intervention, while reliability faults may result from hardware failures and, depending on the implementation, may also arise from external disturbances. From a security perspective, fault injection provides a possible adversarial path to induce a subset of similar behavioral effects. Controlled sensitivity probes vary selected hardware parameters in simulation to isolate their influence on inversion susceptibility. Control in simulation does not imply that a physical adversary can reproduce the same location, magnitude, or repeatability on a fabricated device.

Under these assumptions, we do not claim a chip-specific physical fault-injection exploit. Instead, we characterize whether the model-inversion exposure of a trained model changes after memristive deployment and under different hardware states, including states associated with non-idealities, reliability failures, and hardware faults.

\section{System Architecture}
\label{sec:system}

We describe the baseline \ac{snn} and its mapping to a memristive crossbar architecture. Together, these components define the software, memristive, and reliability fault settings evaluated in this work.

\subsection{Network Topology}

We evaluate compact feedforward convolutional spiking neural networks with Leaky Integrate-and-Fire (LIF) neurons. For the MNIST configuration used in the main analysis, the network contains two $5\times5$ convolutional layers with 16 and 32 channels, respectively, followed by average pooling and a final linear readout. Dataset-specific architectural parameters are summarized in Appendix~\ref{appsec:settings}. Inference is performed over a dataset-specific temporal window; the MNIST configuration uses $T=25$. For static inputs, pixel intensities are converted into spike trains through Poisson rate encoding \cite{neftci2019surrogate, eshraghian2023training}. Native event-based datasets are processed directly through their temporal event representations.

We use the same trained parameters throughout the comparison so that changes in reconstruction behavior can be attributed to hardware mapping and perturbations rather than to differences in training. No explicit privacy defenses are introduced. The selected topology is consistent with edge-oriented neuromorphic inference and crossbar-based implementations; our quantitative results are therefore specific to the evaluated architecture and hardware configuration.

\subsection{Memristive Mapping}
\label{subsec:hwconf}

Synaptic parameters are represented and evaluated through the ReRAM model based on IBM AIHWKIT \cite{le2023using}. Signed synaptic weights $W_{i,j}$ are represented using differential conductance pairs $(G^{+}_{i,j},G^{-}_{i,j})$, with conductance values restricted to $1$--$1000~\mu\mathrm{S}$. The evaluated ReRAM model is parameterized by device-to-device variability with $\sigma_{D2D}=0.1$, cycle-to-cycle noise with magnitude $0.05$, and time-dependent conductance drift. Drift follows the power-law model

\begin{equation}
G(\tau)
=
G(\tau_0)
\left(
\frac{\tau}{\tau_0}
\right)^{-\nu},
\end{equation}

where $\tau$ denotes the modeled time since programming, $\tau_0$ is the reference time, and $\nu$ is the drift coefficient. We use $\nu=0.005$. The crossbar interface additionally uses 8-bit ADC/DAC precision. Table~\ref{tab:architecture} summarizes the evaluated network and hardware configuration.

These parameters define a non-ideal memristive hardware model rather than a transistor-, layout-, or chip-specific reproduction of a fabricated accelerator.

\subsection{Evaluation Settings}
\label{subsec:execution-conditions}

To separate software-level model-inversion behavior from hardware-dependent effects, we evaluate three settings. \textbf{Software} evaluates the trained model $\Theta$ without memristive mapping or analog hardware non-idealities, while preserving the SNN dynamics and input encoding. \textbf{Memristive} maps the same trained parameters to the ReRAM model, including its native hardware non-idealities. \textbf{Memristive and Reliability Fault} additionally evaluates the mapped model under a reliability condition or controlled sensitivity-analysis perturbation $p$.

Using the notation introduced in Section~\ref{subsec:sdmi}, these conditions correspond to $f_{\Theta}^{digital}$, $f_{\Theta,h}^{mem}$, and $f_{\Theta,h,p}^{mem}$, respectively. This comparison separates the effects of memristive mapping from those introduced by additional hardware deviations.

\begin{table}[t]
\small
\centering
\caption{Memristive SNN architecture and evaluated hardware configuration.}
\label{tab:architecture}
\begin{tabularx}{\columnwidth}{lX}
\hline
\textbf{Component} & \textbf{Configuration} \\
\hline

Network
& Compact feedforward CSNN with convolutional layers and a linear readout \\

Neuron Model
& Leaky Integrate-and-Fire (LIF), time-stepped inference with $T=25$ for MNIST \\

Weight Mapping
& Differential ReRAM conductance pairs, $1$--$1000~\mu\mathrm{S}$ \\

Device
& Analog resistive memory with $\sigma_{D2D}=0.1$, C2C noise $=0.05$, and drift $\nu=0.005$ \\

Interface
& 8-bit ADC/DAC \\

Input Encoding
& Poisson rate encoding for static inputs \\

\hline
\end{tabularx}
\end{table}

\section{Reliability Characterization and Sensitivity Analysis}
\label{sec:perturbation-characterization}

Before evaluating the reconstruction leakage, we characterize how hardware non-idealities and reliability faults affect the functional behavior of the target memristive \ac{snn}. Following Section~\ref{subsec:hardware-perturbations}, we consider deviations at synaptic, neuron-threshold, and spike levels, spanning native hardware non-idealities and reliability faults. Figure~\ref{fig:faults} summarizes these abstraction levels, while the complete parameter sweeps are reported in Appendix Fig.~\ref{fig:accuracy_sensitivity}. Their magnitudes are varied through controlled sweeps to characterize regimes from mild to substantial changes in predictive performance.

\subsection{Synaptic-Level Deviations and Weight Sensitivity}

We first consider conductance drift, bit errors, and weight noise. Conductance drift represents a native time-dependent non-ideality of the memristive substrate, whereas bit errors and weight noise model deviations in the effective weight representation.

For conductance drift, we vary the drift coefficient $\nu$ controlling the temporal evolution of the programmed conductance. In the evaluated range, the \ac{snn} remains relatively robust to low drift levels ($\nu < 0.05$), with only limited degradation in classification accuracy, as shown in Appendix Fig.~\ref{fig:accuracy_sensitivity}. This behavior is consistent with the tolerance to moderate analog deviations reported for neural networks mapped to non-ideal hardware \cite{joksas2025nonideality, lammie2025inherent}.

We additionally evaluate discrete bit errors in the weight representation associated with the mapping or peripheral logic. A clear sensitivity transition emerges with increasing bit significance (Appendix Fig.~\ref{fig:accuracy_sensitivity}): errors affecting the most significant bits (MSBs) produce substantially larger accuracy degradation than errors affecting the least significant bits (LSBs), consistent with the larger resulting change in the represented synaptic value \cite{hong2019terminal}. Weight noise provides a complementary stochastic perturbation of the effective synaptic values. Their privacy implications are evaluated separately in Section~\ref{sec:privacy-attacks}.

\subsection{Neuron-Threshold Sensitivity}

We next evaluate deviations affecting the LIF firing threshold $V_{th}$. Threshold deviations are particularly relevant to our privacy analysis because they modify spike activity without directly changing the trained weights. Threshold behavior can be influenced by circuit variability, bias conditions, and changes in the implementation of neurons. We characterize these effects through controlled threshold offsets and stochastic threshold noise.

The sensitivity analysis in Appendix Fig.~\ref{fig:accuracy_sensitivity} reveals a nonlinear relationship between the effective firing threshold and classification accuracy. Moderate reductions in $V_{th}$ do not immediately eliminate predictive functionality in our experiments. Lower thresholds increase neuronal excitability and can lead to higher firing rates and reduced spike sparsity, consistent with prior studies \cite{nagarajan2022analysis, gougebakan2024spikingjet}. Larger deviations eventually produce substantial degradation as the firing dynamics move away from those for which the model was trained.

\subsection{Spike Timing and Transmission Sensitivity}

Finally, we consider spike deletion, insertion, and temporal displacement as behavioral models of missing, spurious, or mistimed events in the spike processing path.

Spike deletion is modeled using a Bernoulli probability $p_{del}$. In the evaluated MNIST configuration, classification accuracy remains above 85\% when 10\% of spike events are randomly removed. The network exhibits stronger functional sensitivity as the temporal displacement $\Delta t$ (number of discrete SNN time steps by which spike events are shifted) increases. These trends are shown in Appendix Fig.~\ref{fig:accuracy_sensitivity}. Unlike random deletion, temporal shifts modify the time at which events contribute to membrane integration and can therefore alter downstream temporal dynamics.

Importantly, functional sensitivity does not imply increased inversion susceptibility. Spike-level perturbations may substantially degrade predictive utility while producing little or no improvement in reconstruction similarity. We therefore evaluate utility degradation and reconstruction leakage as separate quantities in Section~\ref{sec:privacy-attacks}.


\begin{figure}[t]
  \centering
  \includegraphics[width=0.95\linewidth]{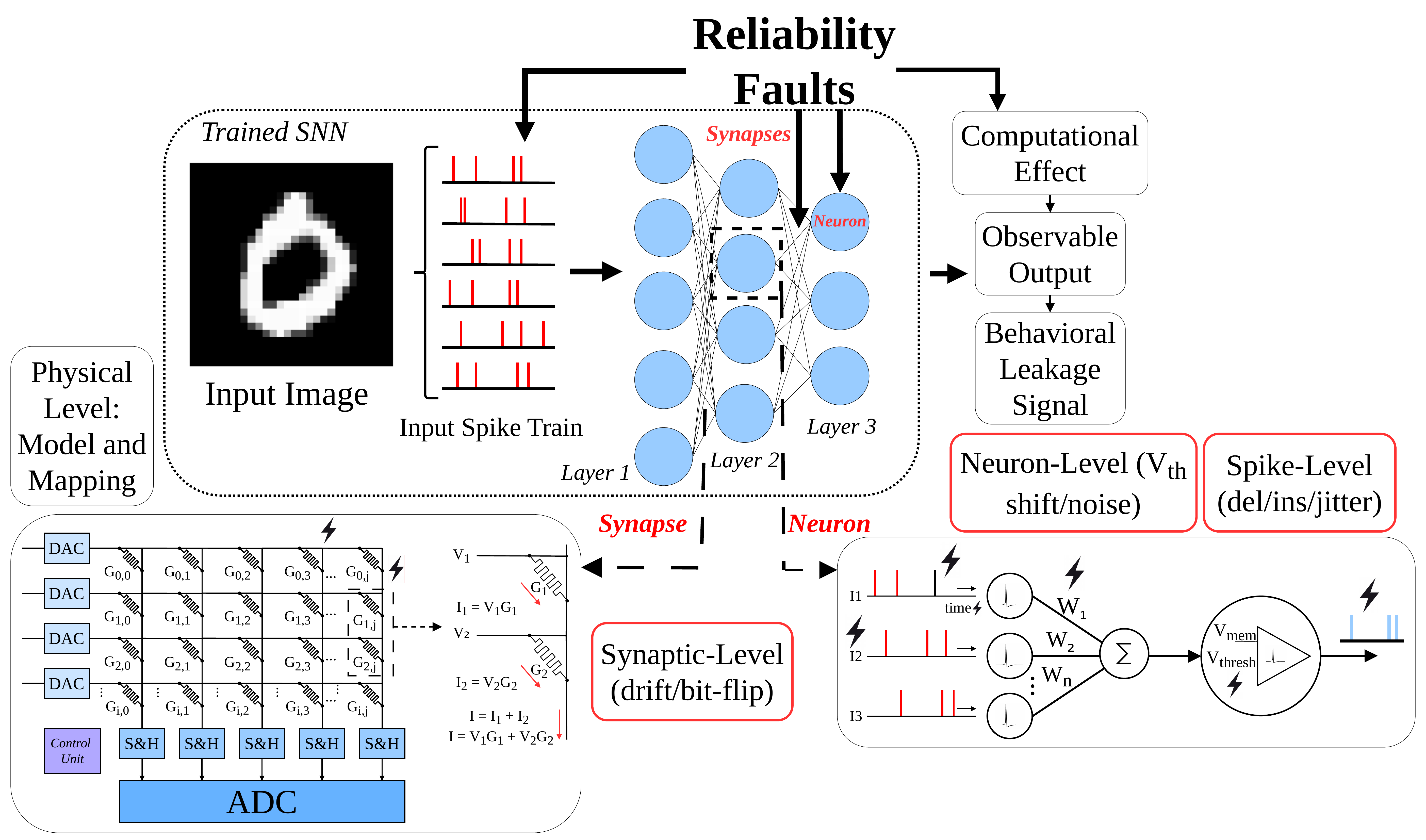}
  \caption{Hardware non-idealities and reliability faults evaluated at synaptic, neuron, and spike levels. Controlled sensitivity sweeps do not imply equivalent physical controllability by an adversary.}
  \label{fig:faults}
\end{figure}

\begin{table}[t]
\centering
\caption{Hardware non-idealities and reliability faults considered in the sensitivity analysis.}
\label{tab:faults}
\begin{tabular}{lll}
\hline
\textbf{Level} & \textbf{Type} & \textbf{Effect} \\
\hline
Synaptic & Drift & Gradual weight change \\
Synaptic & Bit flip & Discrete weight error \\
Synaptic & Weight noise & Induced error in weights \\
Neuron & Threshold offset & Static threshold shift \\
Neuron & Threshold noise & Stochastic threshold shift \\
Spike & Deletion & Missing spike events \\
Spike & Insertion & Spurious spike events \\
Spike & Timing shift & Spike time displacement \\
\hline
\end{tabular}
\end{table}

\section{Spiking Domain Model Inversion (SDMI)}
\label{sec:privacy-attacks}

SDMI extends optimization-based model inversion to temporally encoded \acp{snn} under stochastic memristive execution. We use SDMI as a fixed privacy characterization probe across the software SNN, the memristive SNN, and additional hardware perturbations. Importantly, hardware access is not required to perform inversion on the model; our goal is to determine whether that privacy characterization remains representative after memristive mapping.

\subsection{From Static to Temporal Inversion}

Model Inversion (MI) attacks aim to recover representative inputs or sensitive features associated with a target output of a trained model \cite{wang2021variational}. In optimization-based MI, an input $x$ is iteratively updated so that the model increasingly associates it with a target class $y$.

Applying this approach to \acp{snn} introduces two additional challenges. Inference unfolds over a temporal window $T$, with responses determined by temporally evolving spike activity, and spike generation is non-differentiable. For static inputs, the candidate input $x$ is converted into a spike train $S_x[t]$ through Poisson encoding.

SDMI uses surrogate-gradient estimation through the LIF dynamics \cite{neftci2019surrogate, eshraghian2023training}, together with a straight-through approximation that maps gradients from the sampled spike representation back to the continuous candidate input.

Using the notation introduced earlier, $\Theta$ denotes the trained SNN parameters, $h$ the hardware state, and $p$ the evaluated reliability fault or sensitivity setting. For a target class $y$, SDMI optimizes

\begin{equation}
\begin{aligned}
\min_x \ \mathcal{J}_{p}(x)
=
&\;
\mathbb{E}_{h,S_x}
\left[
\mathcal{L}
\left(
f_{\Theta,h,p}^{mem}(S_x), y
\right)
\right]
\\
&+
\lambda_1 \mathcal{R}_{TV}(x)
+
\lambda_2 \mathcal{R}_{spk}(S_x),
\end{aligned}
\label{eq:sdmi-objective}
\end{equation}

where $S_x \sim \mathcal{P}(x)$ denotes the spike representation generated from the candidate input. $\mathcal{R}_{TV}(x)$ encourages spatial coherence in reconstructed static inputs, while $\mathcal{R}_{spk}(S_x)$ regularizes spike activity during optimization.

The same objective is applied to the three execution conditions defined in Section~\ref{subsec:execution-conditions}: $f_{\Theta}^{digital}$, $f_{\Theta,h}^{mem}$, and $f_{\Theta,h,p}^{mem}$. Keeping the inversion procedure fixed allows changes in reconstruction behavior to be attributed to the execution condition rather than to changes in the attack formulation.

\subsection{Expectation over Transformation}
\label{subsec:eot}

In addition to stochasticity from spike encoding, memristive inference introduces stochasticity from hardware effects. Optimizing against a single sampled hardware state can therefore produce gradients that depend strongly on that hardware state. SDMI uses Expectation over Transformation (EOT) to reduce this dependence \cite{wang2021variational}.

At each optimization step, the gradient of the expected inversion loss with respect to the candidate input $x$ is estimated using $K$ Monte Carlo samples:

\begin{equation}
\label{eq:eot}
\nabla_x
\mathbb{E}_{h,S_x}
\left[
\mathcal{L}
\left(
f_{\Theta,h,p}^{mem}(S_x),y
\right)
\right]
\approx
\frac{1}{K}
\sum_{k=1}^{K}
\nabla_x
\mathcal{L}
\left(
f_{\Theta,h^{(k)},p}^{mem}
\left(
S_x^{(k)}
\right),
y
\right).
\end{equation}

Here, $h^{(k)}$ denotes a sampled hardware state under the evaluated condition $p$, while $S_x^{(k)}$ denotes the corresponding stochastic spike representation. The condition $p$ remains fixed across the $K$ EOT samples; when $p$ defines a stochastic perturbation, its random draw is included in $h^{(k)}$.

Together with surrogate gradients and the straight-through approximation, EOT enables optimization through discrete spiking dynamics while averaging over stochastic spike and hardware samples. Moreover, it reduces dependence on any single sampled execution state rather than assuming knowledge of the instantaneous physical realization. The quantitative effect of EOT on reconstruction similarity, stability, and computational cost is evaluated in Section~\ref{subsec:ablation}.

\subsection{Reconstruction Metrics}
\label{subsec:auxiliary-leakage}

The primary output of SDMI is a reconstructed input representative of a target class. We quantify reconstruction fidelity using Structural Similarity Index (SSIM) and Peak Signal-to-Noise Ratio (PSNR). For each target class, the reconstruction is compared with the nearest same-class test sample, selected by minimum MSE, as detailed in Section~\ref{sec:evaluation}. These metrics characterize class-representative reconstruction fidelity rather than exact recovery of an individual training record.

\begin{figure*}[t]
    \centering
    \includegraphics[width=0.95\linewidth]{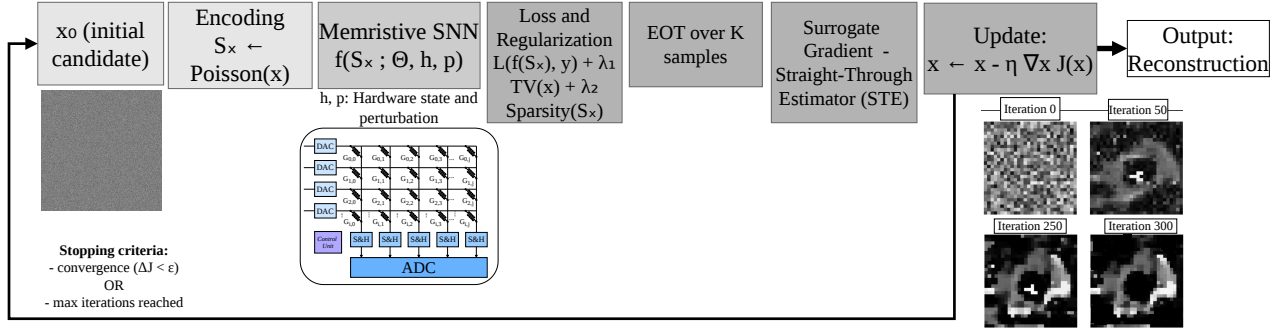}
    \caption{
    Overview of SDMI. A candidate input is converted to its spiking representation and evaluated by the selected SNN model. EOT averages over stochastic spike and hardware samples, while surrogate gradients and the straight-through estimator propagate the inversion objective through the non-differentiable spiking dynamics. The same optimization procedure is applied to the software SNN, the corresponding memristive SNN, and the memristive SNN under additional hardware perturbations.
    }
    \label{fig:sdmi}
\end{figure*}

The overall procedure is illustrated in Fig.~\ref{fig:sdmi}. SDMI generates the candidate spike representation, evaluates the selected hardware condition, estimates the expected inversion loss through EOT, and updates the continuous candidate input using surrogate gradients and a straight-through approximation. The same procedure is applied across all three evaluation conditions.

\section{Evaluation}
\label{sec:evaluation}

This section evaluates how the execution conditions defined in Section~\ref{subsec:execution-conditions} affect model utility and model-inversion leakage. We first compare the SNN and memristive SNNs to isolate the effect of hardware mapping, and then analyze how reliability faults reshape reconstruction behavior and its relationship with predictive utility. Experiments cover MNIST, Fashion-MNIST, AT\&T Faces, and the event-based DVS128 Gesture dataset.

\subsection{Experimental Setup and Metrics}
\label{subsec:experimental-setup}

All networks are trained using surrogate-gradient (SG) learning. The resulting learned parameters are then evaluated across the execution conditions defined in Section~\ref{subsec:execution-conditions}, using the same parameter set for the software and hardware-aware comparisons. Memristive execution uses the ReRAM configuration described in Section~\ref{subsec:hwconf}.

Unless otherwise stated, SNN architectures, LIF configurations, and temporal windows are dataset-specific and are summarized in Appendix~\ref{appsec:settings} (Tables~\ref{tab:settings} and~\ref{tab:hyperparameters}). Static inputs are converted to spike trains using Poisson rate encoding, whereas DVS128 Gesture is processed using its native event representation \cite{eshraghian2023training}. Inversion hyperparameters are kept fixed across execution conditions so that reconstruction differences are not introduced by changes in the SDMI optimization procedure. For reliability characterization, perturbation magnitudes are swept from mild deviations to regimes producing substantial functional degradation. We then select representative operating points to compare perturbation classes at similar utility levels where possible.

We report classification accuracy as the primary utility metric. Reconstruction fidelity is evaluated using Structural Similarity Index (SSIM) and Peak Signal-to-Noise Ratio (PSNR), where applicable. For each target class, SSIM and PSNR are computed against the nearest same-class test sample selected by MSE. These metrics therefore characterize class-representative reconstruction similarity rather than exact recovery of a training record. PSNR provides a complementary pixel-level distortion measure. Target confidence and inversion convergence are treated as auxiliary indicators and are analyzed separately from reconstruction fidelity. For semantically sensitive classes, class-representative reconstruction can nevertheless expose sensitive visual characteristics even without exact recovery of an individual training sample.

For reconstruction experiments, we report the mean and sample
standard deviation across the evaluated target classes. Dataset, training, and hardware parameters are summarized in Tables~\ref{tab:settings} and~\ref{tab:hyperparameters}.

\subsection{SNN and Memristive Baselines}
\label{subsec:digital-memristive}

A central question is whether reconstruction fidelity measured on the software SNN remains representative when the same learned parameters are mapped to non-ideal memristive hardware. We therefore apply the same SDMI configuration to $f_{\Theta}^{digital}$ and $f_{\Theta,h}^{mem}$ while keeping the learned weights and neuron parameters, target class, initialization procedure, inversion budget, and regularization fixed. 

The two execution conditions exhibit nearly identical predictive utility, with accuracies of $96.31\%$ and $95.9\%$ for digital and memristive execution, respectively. For a representative target digit, SDMI also produces similar reconstruction fidelity: SSIM $0.368$ and PSNR $12.38$ dB for the SNN, compared with SSIM $0.360$ and PSNR $12.30$ dB under the memristive neural network.

These results indicate that memristive mapping alone does not account for the larger leakage changes observed under the hardware conditions evaluated below. The following analysis therefore uses memristive SNN as the reference when characterizing how additional hardware deviations affect the reconstruction fidelity.

\subsection{Functional Sensitivity to Reliability Faults}
\label{subsec:functional-sensitivity}

The reliability characterization in Section~\ref{sec:perturbation-characterization} identifies heterogeneous functional responses across synaptic, neuron-threshold, and spike-level deviations. Conductance drift produces gradual changes in effective synaptic efficacy, while discrete bit errors exhibit stronger degradation at higher bit significance \cite{hong2019terminal}. Neuron-threshold and spike-level perturbations likewise produce nonlinear changes in classification performance.

The complete sensitivity sweeps are reported in Appendix Fig.~\ref{fig:accuracy_sensitivity}. We use these sweeps to select representative conditions spanning mild to substantial utility degradation, including conditions with comparable post-perturbation accuracy where possible. These operating regimes provide the basis for the privacy--utility comparisons below; functional degradation alone is not interpreted as evidence of increased or decreased data leakage.

\subsection{Reconstruction Leakage Across Hardware States}
\label{subsec:reconstruction-leakage}

We next evaluate how reliability affects SDMI reconstruction relative to the memristive baseline. As established in Section~\ref{subsec:digital-memristive}, the SNN and its memristive counterpart exhibit comparable baseline utility and reconstruction behavior.

\begin{figure}[t]
  \centering
  \includegraphics[width=0.95\linewidth]{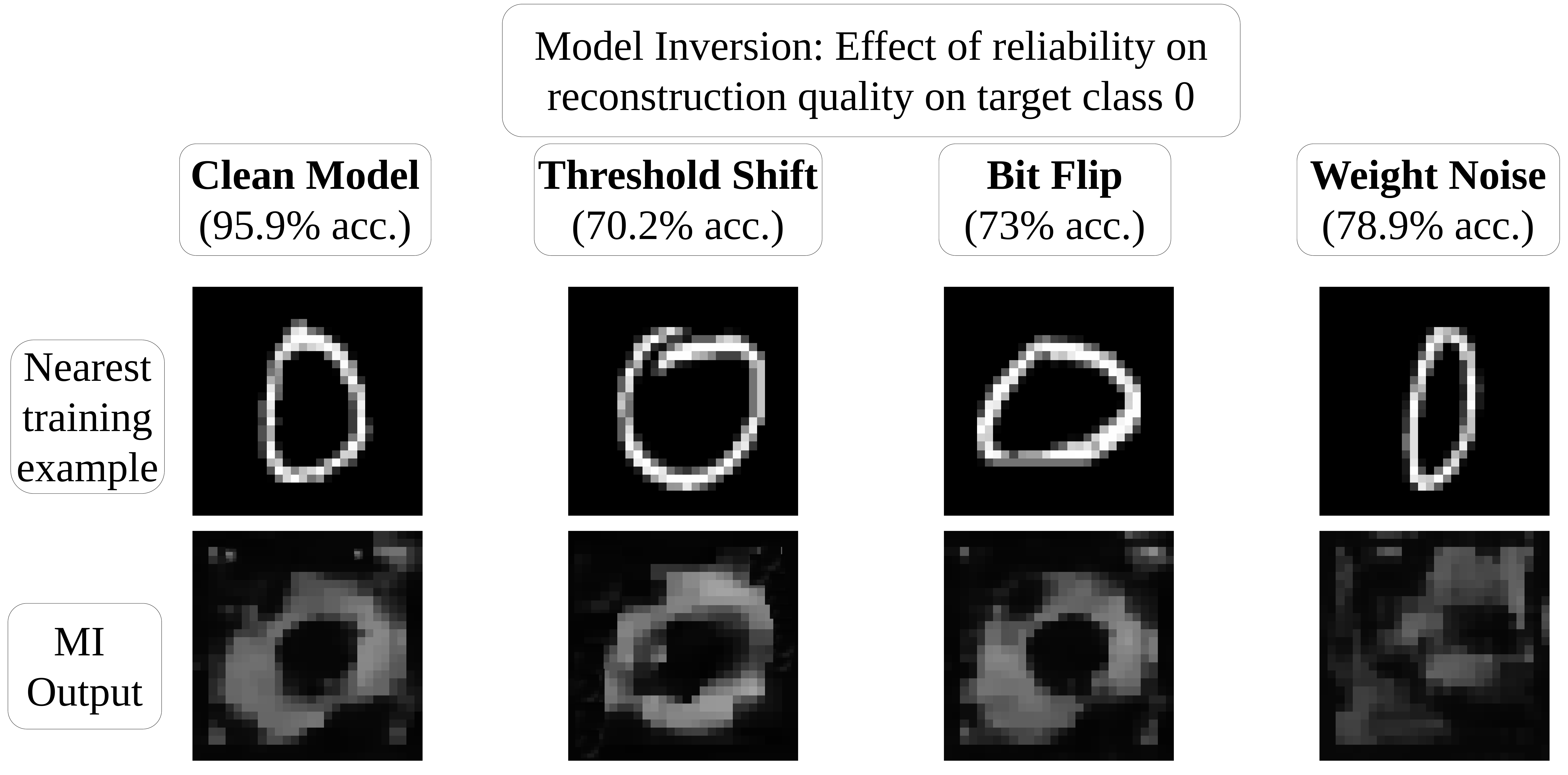}
  \caption{Representative SDMI reconstructions for target class 0 under the memristive baseline and selected hardware perturbations. Synaptic and neuron-threshold perturbations produce distinct changes in reconstruction structure relative to the memristive baseline.}
  \label{fig:reconstruction}
\end{figure}

\begin{figure}[t]
  \centering
  \includegraphics[width=0.95\linewidth]{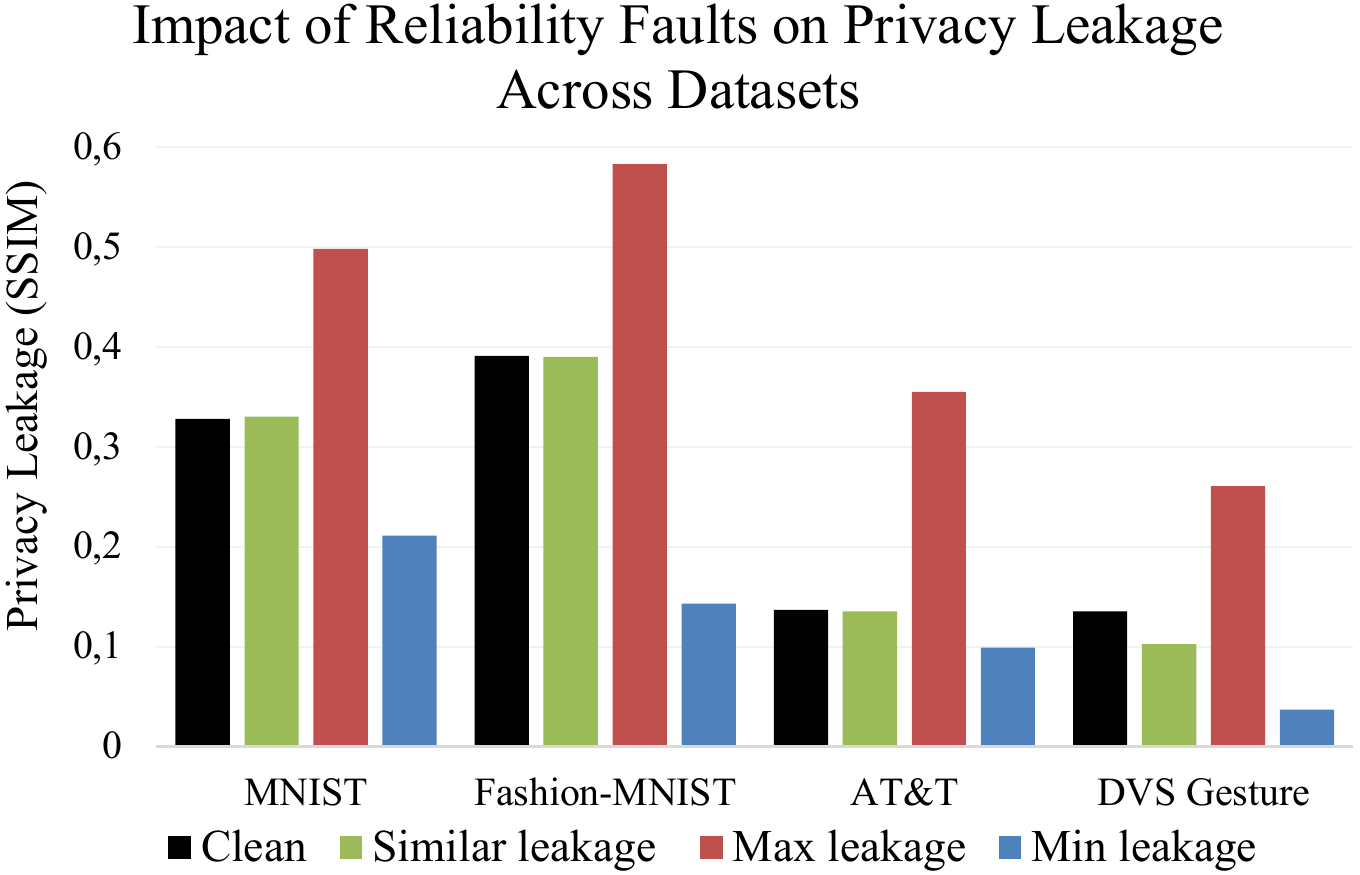}
  \caption{
  Dataset-level comparison of SDMI reconstruction similarity (SSIM) across the evaluated memristive settings. For each dataset, we report the memristive baseline and representative minimum and maximum SSIM values across the evaluated perturbations.
  }
  \label{fig:bar_leakage}
\end{figure}

Figure~\ref{fig:reconstruction} provides representative reconstructions, while Fig.~\ref{fig:bar_leakage} summarizes reconstruction similarity across datasets and hardware conditions.

Under the memristive baseline, SDMI produces structured but imperfect reconstructions. For example, Fashion-MNIST yields an average SSIM of approximately $0.39$ under the reported configuration. These absolute values should be interpreted as reconstruction similarity rather than exact training record recovery.

As summarized in Table~\ref{tab:key_results}, different hardware perturbations affect reconstruction differently. For MNIST, the memristive baseline yields SSIM $0.328$. Conductance drift increases SSIM to $0.351$, while the selected bit-error condition increases it to $0.406$. In contrast, weight noise reduces SSIM to $0.211$.

The strongest increases occur for the evaluated neuron-threshold conditions: threshold shift increases SSIM from $0.328$ to $0.498$, while threshold noise increases it to $0.441$. By contrast, spike deletion, insertion, and temporal displacement remain close to the memristive SSIM baseline despite substantial changes in classification accuracy. Functional sensitivity and reconstruction sensitivity are therefore not equivalent.

\subsection{Reconstruction--Utility Relationship}
\label{subsec:privacy-utility}

We analyze classification accuracy and reconstruction similarity jointly to determine whether functional degradation provides a reliable indication of model-inversion susceptibility. Our results show that it does not: hardware perturbations with comparable post-perturbation accuracy can produce substantially different reconstruction outcomes.

The clearest example is the comparison between neuron-threshold shift and spike deletion as shown in Table~\ref{tab:key_results}. Both reduce classification accuracy to approximately $70.2\%$. However, threshold shift increases SSIM from $0.328$ to $0.498$, whereas spike deletion leaves SSIM approximately unchanged at $0.328$. Comparable utility degradation can therefore correspond to markedly different structural reconstruction leakage.

More broadly, the evaluated perturbations span three qualitatively different regimes. Selected neuron-level conditions increase reconstruction similarity while reducing accuracy; weight noise reduces both accuracy and reconstruction similarity; and spike-level perturbations substantially reduce accuracy while producing little change in SSIM. Neither the magnitude nor direction of the reconstruction change can therefore be inferred from classification accuracy alone.
Accordingly, we do not assign a single scalar notion of reconstruction leakage to all metrics. SSIM and PSNR characterize reconstruction fidelity, and target confidence characterizes optimization toward the target output.

\begin{table}[t]
\centering
\caption{Impact of reliability on utility and SDMI reconstruction similarity.
Accuracy changes ($\Delta$Acc.) are reported in percentage points (pp), and
$\Delta$SSIM denotes the relative percentage change with respect to the
memristive baseline.}
\setlength{\belowcaptionskip}{3pt}
\label{tab:key_results}

\footnotesize
\setlength{\tabcolsep}{3pt}

\begin{tabular}{llcccc}
\hline
\textbf{Level} &
\textbf{Condition} &
\textbf{Acc. (\%)} &
$\Delta$\textbf{Acc. (pp)} &
\textbf{SSIM} &
$\Delta$\textbf{SSIM (\%)} \\
\hline

System &
Memristive &
95.9 &
0.0 &
$0.328 \pm 0.001$ &
0.0 \\

Synaptic &
Drift &
95.6 &
-0.3 &
$0.351 \pm 0.002$ &
+7.0 \\

Synaptic &
Bit error &
73.0 &
-22.9 &
$0.406 \pm 0.003$ &
+23.8 \\

Synaptic &
Weight noise &
78.9 &
-17.0 &
$0.211 \pm 0.004$ &
-35.7 \\

Neuron &
Threshold shift &
70.2 &
-25.7 &
$0.498 \pm 0.004$ &
+51.8 \\

Neuron &
Threshold noise &
81.5 &
-14.4 &
$0.441 \pm 0.004$ &
+34.5 \\

Spike &
Deletion &
70.2 &
-25.7 &
$0.328 \pm 0.001$ &
0.0 \\

Spike &
Insertion &
72.8 &
-23.1 &
$0.328 \pm 0.001$ &
0.0 \\

Spike &
Timing shift &
74.3 &
-21.6 &
$0.329 \pm 0.001$ &
+0.3 \\

\hline
\end{tabular}
\end{table}

\begin{figure}[t]
  \centering
  \includegraphics[width=0.9\linewidth]{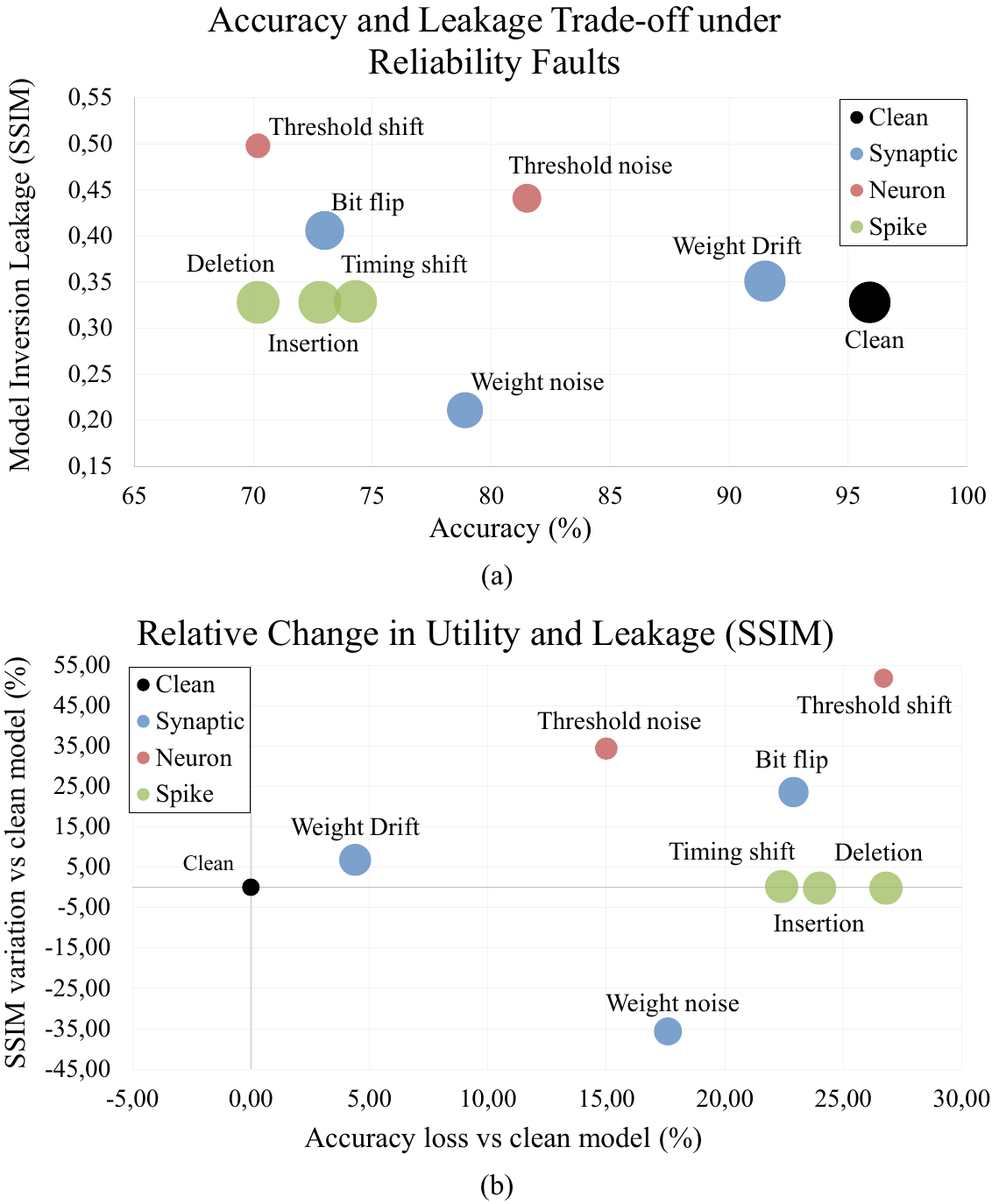}
  \caption{Relationship between classification accuracy and SDMI reconstruction fidelity under the evaluated hardware perturbations: (a) absolute accuracy versus SSIM and (b) relative changes with respect to the memristive baseline. Comparable utility degradation can correspond to substantially different reconstruction behavior; marker size reflects the target confidence.}
  \label{fig:tradeoff}
\end{figure}

The comparison further highlights that perturbation class matters. Neuron-threshold deviations coincide with the largest SSIM increases, whereas spike-level perturbations primarily degrade functional performance without comparable reconstruction gains. Event-based DVS Gesture is more functionally sensitive to temporal displacement than the Poisson-encoded static datasets in our experiments; however, this high functional sensitivity does not by itself imply higher reconstruction fidelity. We therefore interpret these results as hardware-state-dependent behavior rather than as evidence that all perturbations increase model-inversion leakage.

\subsection{Ablation Study: EOT and Regularization}
\label{subsec:ablation}

We finally assess the contribution of EOT and the regularization components of SDMI. Table~\ref{tab:ablation} compares the complete configuration with ablated variants. Removing EOT ($K=1$) decreases mean SSIM from $0.3231$ to $0.2959$ and increases its standard deviation from $0.0012$ to $0.0103$. EOT therefore improves structural reconstruction similarity and, more importantly, stabilizes optimization under hardware and spike stochasticity. PSNR is slightly higher for $K=1$, so we do not interpret EOT as uniformly improving every reconstruction metric.

\begin{table}[t]
\centering
\caption{Ablation study of SDMI components.}
\label{tab:ablation}
\resizebox{\columnwidth}{!}{
\begin{tabular}{lcccc}
\hline
\textbf{Configuration} &
\textbf{SSIM} &
\textbf{PSNR (dB)} &
\textbf{Confidence} &
\textbf{Time (s)} \\
\hline

Full SDMI
& $0.3231 \pm 0.0012$
& $12.5091 \pm 0.0148$
& $0.9995 \pm 0.0000$
& 456.26 \\

No EOT ($K=1$)
& $0.2959 \pm 0.0103$
& $12.9002 \pm 0.0661$
& $0.9978 \pm 0.0003$
& 38.90 \\

No TV
& $0.4014 \pm 0.0703$
& $12.4505 \pm 0.0524$
& $0.9996 \pm 0.0000$
& 455.17 \\

No Sparsity
& $0.3221 \pm 0.0015$
& $12.4872 \pm 0.0242$
& $0.9995 \pm 0.0000$
& 455.90 \\

\hline
\end{tabular}
}
\end{table}

Removing TV increases the mean SSIM to $0.4014$ but also increases its standard deviation substantially, from $0.0012$ to $0.0703$. TV therefore trades maximum mean structural similarity for considerably more consistent reconstruction behavior across evaluated target classes. Removing the sparsity regularizer has only a minor quantitative effect, suggesting that it acts as a secondary structural prior in this configuration. We additionally vary the number of EOT samples $K$ in Table~\ref{tab:eot}. Increasing $K$ monotonically improves mean SSIM over the evaluated range, while runtime grows approximately linearly, exposing a direct trade-off between Monte Carlo averaging and inversion cost.

\begin{table}[t]
\centering
\caption{Sensitivity of SDMI to the number of EOT samples $K$.}
\label{tab:eot}
\begin{tabular}{lccc}
\hline
\textbf{$K$} &
\textbf{SSIM} &
\textbf{Confidence} &
\textbf{Time (s)} \\
\hline
1  & $0.2989 \pm 0.0051$ & $0.9978 \pm 0.0002$ & 39.05 \\
4  & $0.3130 \pm 0.0035$ & $0.9993 \pm 0.0001$ & 152.27 \\
8  & $0.3192 \pm 0.0026$ & $0.9995 \pm 0.0001$ & 304.53 \\
12 & $0.3223 \pm 0.0018$ & $0.9995 \pm 0.0001$ & 455.15 \\
20 & $0.3259 \pm 0.0019$ & $0.9996 \pm 0.0001$ & 758.36 \\
\hline
\end{tabular}
\end{table}

\section{Discussion}
\label{sec:discussion}

\paragraph{Reconstruction does not track model utility.}
Our results challenge the assumption that model-inversion leakage varies monotonically with model accuracy. In memristive \acp{snn}, reconstruction leakage does not follow a monotonic relationship with inference accuracy. Instead, it depends on the interaction between temporal neural dynamics and the effective hardware execution state, indicating that model-inversion susceptibility should be assessed at the system level rather than inferred solely from model behavior. In particular, spike deletion substantially degrades classification accuracy while leaving SSIM approximately unchanged, whereas neuron-threshold perturbations can produce comparable utility degradation together with substantially higher reconstruction similarity. This behavior directly corresponds to the trade-off regions observed in Fig.~\ref{fig:tradeoff}.

\paragraph{Hardware deviations reshape inversion.}
A key observation is that selected hardware deviations can substantially reshape reconstruction leakage. This effect is most evident in neuron-level and selected synaptic perturbations, where changes in threshold dynamics or weight representations coincide with stronger reconstruction structure for the target class. Importantly, this behavior is not universal: weight noise reduces reconstruction similarity, while spike-level perturbations largely degrade utility without comparable SSIM gains. Reliability and fault-injection studies in neural accelerators and memristive systems have primarily focused on functional degradation rather than information exposure \cite{hong2019terminal, sapui2024power, staudigl2022x}. Multiple perturbation sources may also co-occur in practice, and understanding their combined effect on reconstruction leakage remains an important direction for future work.


\paragraph{Similar utility can hide different leakage.}
The comparison between spike deletion and threshold shift further illustrates this privacy--utility decoupling. Both reduce accuracy to approximately $70\%$, yet spike deletion remains close to the memristive SSIM baseline, whereas threshold shift increases SSIM by more than $50\%$. Conversely, weight noise reduces inversion fidelity in our evaluated setting, showing that hardware variability is not inherently privacy-degrading. The reduction observed under weight noise suggests that controlled noise injection could be explored as a privacy-enhancing mechanism. 

\paragraph{Temporal dynamics amplify hardware sensitivity.}
The temporal dimension of spiking computation provides an additional mechanism through which hardware deviations can affect model behavior. Because inference is performed over multiple time steps, deviations can accumulate and propagate through layers by modifying firing activity, threshold crossings, and spike timing. Such hardware-induced changes are not captured by analyses that consider only the trained model. Previous studies on SNN robustness and fault tolerance similarly report sensitivity to neuron-, synapse-, and spike-level perturbations, although they do not study their consequences for model-inversion leakage \cite{jouni2025stdp}. Our results extend this perspective by showing that functional sensitivity and reconstruction sensitivity need not evolve together.

\paragraph{Hardware state matters for privacy assessment.}
More generally, evaluating only the model or post-deployment task accuracy may overlook changes in model-inversion leakage introduced by the hardware substrate. The behavior of a memristive model depends on device- and circuit-level characteristics such as conductance variability, drift, and peripheral precision, and different execution states may therefore exhibit different leakage profiles even at similar utility levels. Variations across devices, fabrication conditions, or operational states could plausibly contribute to such differences, although quantifying device-to-device privacy variation requires physical measurements beyond the scope of our evaluation \cite{zhao2019memory, huang2023hardware, raptis2025input}. For device and materials researchers, these results suggest that hardware variability may have implications beyond functional accuracy and reliability: device-level properties such as conductance variability and drift can also influence the privacy profile of the mapped model. This motivates considering privacy-related behavior alongside conventional device and system metrics when evaluating emerging neuromorphic hardware.

These observations motivate treating hardware state as an explicit dimension of privacy assessment in computation-in-memory and neuromorphic systems. A model characterized under one execution condition may exhibit a different leakage profile under another condition with distinct variability or reliability characteristics \cite{wang2024cimsat, goswami2023experimental}. This consideration is particularly relevant for edge-oriented neuromorphic systems, where inference increasingly occurs on specialized local hardware \cite{liu2023neuropots, cherupally2021leveraging}. Privacy analysis should therefore distinguish the model, its hardware-mapped implementation, and relevant reliability states rather than assuming that privacy properties measured on the software model remain invariant after deployment.

\paragraph{Limitations.}
Our study provides a simulation-based characterization of model-inversion susceptibility in compact memristive SNNs and does not constitute physical-chip validation or demonstrate a chip-specific fault-injection mechanism. The quantitative results are specific to the evaluated SNN architectures, datasets, modeled ReRAM configuration, and perturbation ranges. In addition, SSIM and PSNR quantify class-representative reconstruction fidelity and do not establish exact recovery of individual training records. Extending the analysis to larger and more diverse SNN architectures, multiple device technologies and hardware realizations, independent semantic leakage metrics, and measured device characteristics remains important future work.

More broadly, our results indicate that privacy analysis of memristive neuromorphic systems should not be restricted to the software-level model. This view aligns with emerging work on secure compute-in-memory architectures and broader efforts in machine-learning security to account for hardware and system effects \cite{wang2024safe, bi2023ad2vncs, rajabi2023mdtd, maia2022can}. Extending this analysis to on-chip learning, larger neuromorphic architectures, and emerging memory technologies represents an important step toward a more complete understanding of privacy and security in neuromorphic systems.

\section{Conclusion}

This paper studies model-inversion leakage in memristive spiking neural networks under non-ideal memristive hardware behavior and reliability faults. Using Spiking Domain Model Inversion (SDMI), we characterize how reconstruction behavior changes across execution conditions and show that model utility and reconstruction leakage do not necessarily evolve together. In particular, selected threshold perturbations increase reconstruction similarity by up to 51.8\% relative to the memristive baseline, whereas weight noise reduces it by 35.7\%. Moreover, spike deletion can produce comparable accuracy degradation to threshold shift while leaving reconstruction similarity close to the baseline.

These results show that hardware deviations can reshape the model-inversion leakage profile of a deployed \ac{snn} rather than simply introducing uniform degradation or noise. Classification accuracy alone is therefore insufficient to characterize model-inversion exposure. More broadly, our findings motivate privacy evaluations that differentiate the model, its hardware-mapped implementation, and relevant reliability states. For memristive \acp{snn}, privacy should therefore be assessed as a property of the deployed execution system rather than assumed to remain invariant from the software model.

\section*{Acknowledgments}
This work was partially supported by the European Union through the Horizon Europe Marie Skłodowska-Curie Grant Agreement No 101207193, and the European Research Council (ERC) through the European’s Union Horizon Europe Research and Innovation Programme under Grant Agreement No 101042585. Views and opinions expressed are however those of the authors only and do not necessarily reflect those of the European Union, the European Research Council or the European Research Executive Agency. Neither the European Union nor the granting authority can be held responsible for them.

\bibliographystyle{ACM-Reference-Format}
\bibliography{references}

\appendix

\section{Experimental Settings}
\label{appsec:settings}


\begin{table}[!htbp]
\centering
\caption{Overview of the experimental evaluation protocol.}
\label{tab:settings2}
\small
\setlength{\tabcolsep}{3pt}
\renewcommand{\arraystretch}{1.05}

\begin{tabularx}{\columnwidth}{
@{}
>{\raggedright\arraybackslash}p{0.27\columnwidth}
>{\raggedright\arraybackslash}X
@{}
}
\hline
\textbf{Item} & \textbf{Setting} \\
\hline

Datasets &
MNIST, Fashion-MNIST, AT\&T Faces, DVS128 Gesture \\

Encoding &
Poisson (static), event-based (DVS) \\

Temporal window &
$T=25$ (static), $T=50$ (DVS) \\

Training &
Surrogate-gradient learning \\

Evaluation &
Software, memristive, and reliability-fault conditions \\

Perturbations &
Synaptic, neuron, and spike levels \\

Aggregation &
Mean $\pm$ sample std. across target classes \\

\hline
\end{tabularx}
\end{table}



\begin{table}[!htbp]
\centering
\caption{Dataset-specific SNN configurations used in the evaluation.}
\label{tab:settings}
\scriptsize
\setlength{\tabcolsep}{2pt}
\renewcommand{\arraystretch}{1.06}

\begin{tabularx}{\columnwidth}{
@{}
>{\raggedright\arraybackslash}p{0.22\columnwidth}
>{\raggedright\arraybackslash}X
@{}
}
\toprule
\textbf{Dataset} & \textbf{Configuration} \\
\midrule

MNIST &
10 classes; Poisson; $T=25$.
Conv(1,16,$5{\times}5$)--LIF--AvgPool;
Conv(16,32,$5{\times}5$)--LIF--AvgPool;
FC(10). \\

Fashion-MNIST &
10 classes; Poisson; $T=25$.
Conv(1,16,$5{\times}5$)--LIF--AvgPool;
Conv(16,32,$5{\times}5$)--LIF--AvgPool;
FC(10). \\

AT\&T Faces &
40 classes; Poisson; $T=25$.
Conv(1,32,$5{\times}5$)--LIF--AvgPool;
Conv(32,64,$5{\times}5$)--LIF--AvgPool;
Conv(64,128,$3{\times}3$)--LIF--AvgPool;
FC(40). \\

DVS128 Gesture &
11 classes; event-based; $T=50$.
Conv(2,32,$5{\times}5$)--LIF--AvgPool;
Conv(32,64,$5{\times}5$)--LIF--AvgPool;
FC(11). \\

\bottomrule
\end{tabularx}
\end{table}



\begin{table}[!htbp]
\centering
\caption{Hyperparameters and hardware-aware configuration.}
\label{tab:hyperparameters}
\small
\setlength{\tabcolsep}{2pt}
\renewcommand{\arraystretch}{1.05}

\begin{tabularx}{0.96\columnwidth}{
@{}
>{\raggedright\arraybackslash}p{0.29\columnwidth}
>{\raggedright\arraybackslash}X
@{}
}
\toprule
\textbf{Category} & \textbf{Parameter and Value} \\
\midrule

\textbf{SNN (LIF)}
& LIF parameters: dataset/layer-specific \\

&
Reset potential ($V_{\mathrm{reset}}$): 0.0 \\

&
Time steps ($T$): dataset-specific ($T=25,50$) \\

\midrule

\textbf{IBM AIHWKIT}
& Device: ReRAM \\

&
Conductance range ($G$): $[1,1000]~\mu\mathrm{S}$ \\

&
D2D variability ($\sigma_{\mathrm{D2D}}$): 0.1 \\

&
Cycle-to-cycle noise: 0.05 \\

&
Drift coefficient ($\nu$): 0.005 \\

&
ADC/DAC precision: 8-bit \\

\midrule

\textbf{SDMI Attack}
& Optimizer: Adam \\

&
Initial learning rate ($\eta$): 0.01 \\

&
EOT samples ($K$): 12 \\

&
Iterations: 1200 \\

&
Regularization: two-stage schedule
(Table~\ref{tab:sdmi_params}) \\

\midrule

\textbf{Hardware}
& GPU: NVIDIA GeForce RTX 5070 \\

&
Precision: mixed FP16/FP32 \\

\bottomrule
\end{tabularx}
\end{table}



\begin{table}[!htbp]
\centering
\caption{SDMI optimization stages and regularization schedule.}
\label{tab:sdmi_params}
\small
\setlength{\tabcolsep}{2pt}
\renewcommand{\arraystretch}{1.05}

\begin{tabular}{
@{}
p{0.48\columnwidth}
>{\centering\arraybackslash}p{0.17\columnwidth}
>{\centering\arraybackslash}p{0.17\columnwidth}
@{}
}
\hline
\textbf{Parameter} & \textbf{Stage 1} & \textbf{Stage 2} \\
\hline

Learning rate
& 0.01
& 0.003 \\

Start iteration
& 0
& 250 \\

Sparsity weight
& 0.5
& 8.0 \\

FR excess weight
& 0.5
& 3.0 \\

FR $\ell_2$ weight
& 0.5
& 6.0 \\

TV weight
& 0.03
& 0.18 \\

Image $\ell_2$ weight
& 0.05
& 0.15 \\

Blur weight
& 0.015
& 0.20 \\

Mean-intensity weight
& 0.8
& 1.2 \\

Border weight
& 0.5
& 1.0 \\

\hline
\end{tabular}
\end{table}


Adam is used for 1200 iterations with margin weight $1.0$,
firing-rate target $0.02$, base rate $1.0$, blur kernel $3$,
gradient clipping $1.0$, and initialization $z=-2.2$ with
$x=\sigma(z)$.

\paragraph{Temporal dynamics.}
Inference is performed over discrete time steps ($T=25,50$), where
information is represented through spike timing and firing activity.
This temporal representation is central to our analysis because both
inference and SDMI operate over temporally evolving spike responses
rather than conventional static activations.

\paragraph{Hardware-aware inference.}
The trained SNNs are evaluated in software and after mapping to the
IBM AIHWKIT analog in-memory computing model. The memristive model
incorporates the hardware non-idealities specified in
Table~\ref{tab:hyperparameters}, including device-to-device
variability, cycle-to-cycle noise, conductance drift, and
finite-precision peripheral conversion. This provides a controlled
hardware-aware simulation environment for studying how non-ideal
analog behavior affects model behavior; it should not be interpreted
as a chip-level reproduction of a specific fabricated accelerator.

\paragraph{Execution conditions.}
As defined in Section~\ref{subsec:execution-conditions}, we define
three execution conditions throughout the evaluation:
(i) software execution,
(ii) memristive execution under the modeled hardware non-idealities,
and
(iii) memristive execution under additional reliability faults or
sensitivity-analysis conditions.
The trained model parameters and SDMI configuration are kept fixed
when comparing these conditions.

\paragraph{Attack optimization.}
The SDMI attack operates in the temporal domain by optimizing
candidate inputs over spike-based model responses. Surrogate gradients
provide a differentiable approximation of the spike-generation
function, while Expectation over Transformation (EOT) averages the
inversion objective over stochastic input and hardware execution
realizations.

\paragraph{Reconstruction metrics.}
Inputs are normalized to $[0,1]$. For each target class,
reconstruction metrics are computed against the nearest same-class
test example among up to 200 candidates, selected by minimum MSE.
SSIM is computed from full-image mean, variance, and covariance using
$C_1=(0.01)^2$ and $C_2=(0.03)^2$; PSNR assumes a unit data range.
Reported mean and sample standard deviation values aggregate results
across target classes.

\paragraph{Comparison protocol.}
To preserve comparability across execution conditions, the learned model parameters and SDMI configuration are not retuned after switching from software execution to the memristive or reliability-perturbed settings. The inversion budget, regularization schedule, initialization procedure, and target-class protocol therefore remain fixed, while only the evaluated hardware execution state changes. This design isolates hardware-dependent changes in reconstruction behavior from changes introduced by model retraining or attack-specific hyperparameter tuning.

\paragraph{Interpretation of the sensitivity sweeps.}
The sensitivity curves are intended to characterize how different classes of hardware deviations affect functional behavior over increasing perturbation magnitude. Because the parameters associated with synaptic, neuron, and spike-level deviations have different physical meanings and scales, equal numerical perturbation values are not interpreted as equivalent hardware severity. Instead, the sweeps are used to identify representative operating regimes and, where possible, conditions with comparable post-perturbation utility. The selected points are therefore characterization points rather than optimized attack parameters.



\section{Additional Perturbation Sensitivity Results}
\label{app:perturbation-sensitivity}

\begin{figure}[H]
    \centering
    \includegraphics[width=0.9999\columnwidth]{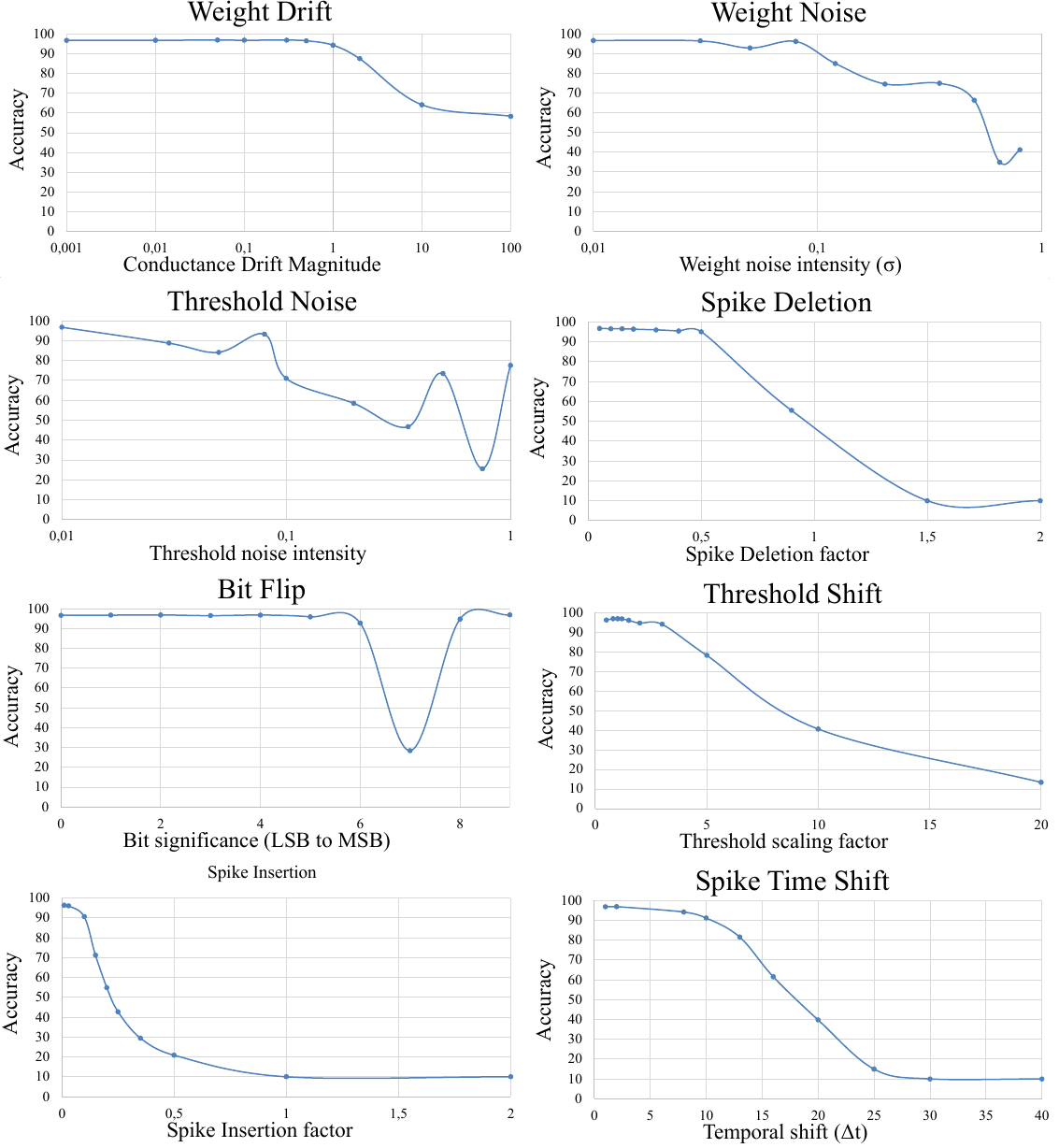}
    \caption{Accuracy sensitivity to perturbation magnitude for the evaluatedhardware conditions. 
    }
    \label{fig:accuracy_sensitivity}
\end{figure}

We provide additional classification-accuracy sensitivity curves for
MNIST for all considered hardware and reliability faults as a
function of perturbation magnitude in Fig.~\ref{fig:accuracy_sensitivity}.
Table~\ref{tab:selected_perturbations} summarizes the representative
operating points used in the privacy analysis. These sweeps characterize
functional sensitivity and are used to identify the operating regimes
analyzed in the main paper; they do not imply equivalent physical
controllability of the corresponding perturbations.

\begin{table}[H]
\centering
\caption{Representative perturbation settings used in the privacy analysis.}
\label{tab:selected_perturbations}
\small
\setlength{\tabcolsep}{3pt}

\begin{tabular}{lll}
\hline
\textbf{Condition} & \textbf{Parameter} & \textbf{Selected value} \\
\hline
Weight drift      & Multiplicative deviation $\delta$ & $1$ \\
Bit error         & Bit position $b$                   & $6$ \\
Weight noise      & Std. dev. $\sigma_w$               & $0.56$ \\
Threshold shift   & Threshold scale                    & $10$ \\
Threshold noise   & Std. dev. $\sigma_{th}$            & $0.07$ \\
Spike deletion    & Probability $p_{\mathrm{del}}$     & $0.75$ \\
Spike insertion   & Probability $p_{\mathrm{ins}}$     & $0.21$ \\
Timing shift      & Displacement $\Delta t$            & $13$ steps \\
\hline
\end{tabular}
\end{table}

\section{Ethical Considerations}

This work studies privacy risks in neuromorphic hardware under
hardware non-idealities and reliability faults. Although such analysis
has dual-use implications, our goal is defensive: to identify
hardware-dependent privacy leakage mechanisms and motivate more secure
evaluation and hardware--software co-design practices.

All experiments are conducted in simulation using public benchmark
datasets. The work does not involve human subjects, private user data,
interaction with deployed systems, or physical fault injection on
operational hardware. The presented results are intended to support
hardware-aware privacy assessment rather than operational misuse.



% ...

\end{document}